\documentclass[a4paper,11pt]{article}
\usepackage{pos}

\title{A phenomenological note on the missing $\rho_2$ meson}

\author[]{Shahriyar Jafarzade}

\affiliation[]{Institute of Physics, Jan Kochanowski University, \\
  ul. Uniwersytecka 7, 25-406, Kielce, Poland,}

\emailAdd{shahriyar.jzade@gmail.com}

\abstract{The $\rho_2$ meson is the missing isovector member of the meson nonet with the quantum numbers $J^{PC}=2^{--}$. It belongs to the class of $\rho$-mesons such as the vector meson $\rho(770)$, the excited vector  $\rho(1700)$ and the tensor $\rho_3(1690)$. Yet, despite the rich experimental and theoretical studies for other $\rho$-meson states, no resonance that could be assigned to the $\rho_2$ meson has been measured. In this note, we present the results for the mass and dominant decay channels of the $\rho_2$ meson within the extended Linear Sigma Model. }

\FullConference{%
  41st International Conference on High Energy Physics - ICHEP2022\\
  6-13 July, 2022\\
  Bologna, Italy
}

\begin{document}
\maketitle

\section{Introduction}
PDG contains various mesons denoted with the letter $\rho$ \cite{Workman:2022ynf} . These are the isovector resonances with quantum number of isospin ($I=1$), of parity $(P=-1)$, and of charge conjugation ($C=-1$). For instance, the vector mesons $\rho(770)$ with  quantum number $J^{PC}=1^{--}$ \cite{Parganlija:2012fy}, the  excited vector mesons $\rho(1450)$, $\rho(1700)$ \cite{Piotrowska:2017rgt,Piotrowska:2017idc,Piotrowska:2017yfu}, and the tensor meson $\rho_3(1690)$ with quantum number $J^{PC}=3^{--}$ \cite{Jafarzade:2021vhh}.  Despite the prediction of the $\rho_2$ in the Relativistic Quark model \cite{Godfrey:1985xj},  it is still missing experimentally. We only  have the following data which were observed from different experimental groups and listed as ``further states'' in PDG \cite{Workman:2022ynf}: $ \rho_2(1940)$ and $ \rho_2(2225)$ with the total decay widths  $\Gamma^{\text{tot}}_{\rho_2(1940)} \simeq 155 \pm 40 $ MeV and   $\Gamma^{\text{tot}}_{\rho_2(2225)} \simeq 335^{+100}_{-50}$ MeV accordingly.   Axial tensor mesons are studied in  recent LQCD simulations \cite{Johnson:2020ilc}, where the authors consider the mass of $\rho_2$ is about $1.7$ GeV as the $\rho_3(1690)$. We present the results about the  missing $\rho_2$ \cite{Jafarzade:2022uqo}  within a chiral effective model which is so-called the extended Linear Sigma Model (eLSM) \cite{Parganlija:2012fy}.

\section{Effective Model and Results}
The physical resonances such as the pseudoscalar mesons  \big\{$\pi, K, \eta, \eta^\prime(958)$\big\}, the vector mesons  \big\{$\rho(770)$, $\overline{K}^\star(892)$, $\omega(782)$, $\phi(1020)$\big\}, and the tensor mesons  \big\{$a_2(1320)$, $\overline{K}^\star_2(1430)$, $f_2(1270)$, $f_2^\prime(1525)$\big\} together with their chiral partners construct chiral nonets.
Table \ref{tab:chiral-transformations} describes the transformation of the chiral fields under different symmetries for (pseudo) scalars (P) S, (axial-) vector ($A_1^\mu$) $V^{\mu}$ and (axial-) tensor ($A_{2}^{\mu\nu}$) $T^{\mu\nu}$ nonets.
	\begin{table}[h]
   	\centering
		\renewcommand{\arraystretch}{1.1}
		\begin{tabular}[c]{|c|c|c|c|}
			\hline
			Nonet & Parity $(P)$ & Charge conjugation $(C)$ & $U_R(3) \times U_L(3)$ \\
			\hline \hline
			$\Phi(t,\vec{x}):=S(t,\vec{x})+iP(t,\vec{x})$ & $\Phi^{\dagger}(t,-\vec{x})$ & $\Phi^{t}(t,\vec{x})$ & $U_L\Phi U^{\dagger}_R$ \\
			\hline
			$R^{\mu}(t,\vec{x}):=V^{\mu}(t,\vec{x})-A_1^{\mu}(t,\vec{x})$ & $L_{\mu}(t,-\vec{x})$ & $-(L^{\mu }(t,\vec{x}))^{t}$ & $U_R R^{\mu}U_R^{\dagger}$ \\
			\hline
			$L^{\mu}(t,\vec{x}):=V^{\mu}(t,\vec{x})+A_1^{\mu}(t,\vec{x})$ & $R_{\mu}(t,-\vec{x})$ & $-(R^{\mu }(t,\vec{x}))^{t}$ & $U_L L^{\mu}U_L^{\dagger}$ \\
			\hline
			$\mathbf{R}^{\mu\nu}(t,\vec{x}):=T^{\mu \nu}(t,\vec{x})-A_2^{\mu \nu}(t,\vec{x})$ & $\mathbf{L}_{\mu\nu}(t,-\vec{x})$ & $(\mathbf{L}^{\mu\nu}(t,\vec{x}))^{t}$ & $U_{R}\mathbf{R}^{\mu\nu}U^{\dagger}_{R}$ \\
			\hline
			$\mathbf{L}^{\mu\nu}(t,\vec{x}):=T^{\mu \nu}(t,\vec{x})+A_2^{\mu \nu}(t,\vec{x})$ & $\mathbf{R}_{\mu\nu}(t,-\vec{x})$ & $(\mathbf{R}^{\mu\nu}(t,\vec{x}))^{t}$ & $U_{L}\mathbf{L}^{\mu\nu}U^{\dagger}_{L}$ \\
			\hline
		\end{tabular}
		\caption{Transformations of the chiral multiplets under P, C, and $U_R(3) \times U_L(3)$.}\label{tab:chiral-transformations}
					\end{table}

The chiral invariant Lagrangian that generates the masses of the spin-2 mesons reads
\begin{align}\label{eq:mLag}
    \mathcal{L}_{\text{mass}}=\text{Tr}\Big[\Big(\frac{m_{\text{ten}}^{2}}{2}+\Delta^{\text{ten}}\Big)\Big(\mathbf{L}_{\mu\nu}^2+\mathbf{R}_{\mu\nu}^2\Big)
  + \mathbf{R}^{\mu\nu} \mathbf{R}_{\mu\nu}\Big] +2 h_3^{\text{ten}}\text{Tr}\Big[\Phi \mathbf{R}^{\mu\nu}\Phi^\dagger \mathbf{L}_{\mu\nu} \Big]
   \text{,}
\end{align}
where $\Delta^{\text{ten}}=\text{diag}\{0\,,0\,,\delta_{S}^{\text{ten}}= m_{K_{2}}^2-m_{\textbf{a}_2}^2\}$.

The following three equations are coming from the extended version of the above lagrangian and relate the masses of spin-2 chiral partners:
\begin{align}
m_{\rho_2}^2= m_{a_2}^2- h_3^{\text{ten}}\phi_N^2,\quad
m_{K_{2A}}^2= m_{K_{2}}^2-\sqrt{2} h_3^{\text{ten}}\phi_N\phi_S,\quad
m_{\omega_{2,S}}^2&=m_{f_{2,S}}^2-2 h_3^{\text{ten}}\phi_S^2 \text{ ,}
\end{align}
which leads to $m_{\rho_2}=1663$ MeV where we have assumed $m_{K_{2A}}=m_{K_{2}(1820)}$. The same assumption in the last term of Eq (\ref{eq:mLag}) implies $\Gamma(\rho_2\rightarrow a_2(1320))\pi\approx 88$ MeV which is 200 MeV in \cite{Guo:2019wpx}. Our prediction for the mass of $\rho_2$ from the spontaneous breaking of the chiral symmetry is near to the  prediction in \cite{Godfrey:1985xj}.

The simplest Lagrangian which describes tree level decays has the following form
\begin{align}\label{lag1}
    \mathcal{L}=\frac{g_2^{\text{ten}}}{2}\Big(\text{Tr}\Big[ \mathbf{L}_{\mu\nu}\{ L^{\mu}, L^{\nu}\}\Big]+\text{Tr}\Big[\mathbf{R}_{\mu\nu} \{R^{\mu}, R^{\nu}\} \Big]\Big) 
    \text{ .}
\end{align}
We firstly present the results for $a_2(1320)$ with the quantum number $J^{PC}=2^{++}$  based on the Lagrangian (\ref{lag1}).
\begin{table}[h]
\centering
\renewcommand{\arraystretch}{1.1} 
\begin{tabular}{|l|c|c|}
\hline
Decay process (in model)  & eLSM \cite{Jafarzade:2022uqo}     &  PDG \cite{Workman:2022ynf}  \\ \hline\hline
	$\,\;a_2(1320) \longrightarrow  \bar{K}\, K  $ &  $4.06\pm 0.14$  & $7.0^{+2.0}_{-1.5}$
\\ \hline
	$\,\;a_2(1320) \longrightarrow \pi\,\eta   $  &  $25.37\pm 0.87$   & $18.5\pm 3.0$ 
\\ \hline
$\,\;a_2(1320) \longrightarrow  \pi\,\eta^\prime (958)  $ &  $1.01\pm 0.03$   & $0.58\pm 0.10$
\\ \hline
\end{tabular}%
\caption{Decay rates of the $a_2(1320)$ into the pseudoscalar mesons  in MeV.
}\label{tabrtpp}
\end{table}
Secondly, we present the results in Table \ref{tab:rho2} for the missing $\rho_2$. Note that, we have used the PDG data in Table \ref{tabrtpp} to obtain the coupling $g_2^{\text{ten}}$ for presenting the results in the second row of the Table \ref{tab:rho2}. We expect the dominant $\rho_2$ decay widths in the interval between the second and the third rows of the following table which implies it is being broad despite some uncertainties.
\begin{table}[h] 
		\centering
		\renewcommand{\arraystretch}{1.1}
		\begin{tabular}[c]{|c|c|c|c|}
			\hline
			Decay process (in model) & eLSM  &  eLSM (LQCD)   & LQCD   \cite{Johnson:2020ilc}  \\
			\hline \hline
			$\,\;\rho_2(?) \longrightarrow \rho(770)\, \eta$ & $87$ & $30$  & $-$  \\
			\hline
			$\,\;\rho_2(?) \longrightarrow \bar{K}^\ast(892)\, K +\textbf{c}.\mathrm{c}.$ & $ 77$  & $27$  & $36$ \\
			\hline
		$\,\;\rho_2(?) \longrightarrow \omega(782)\, \pi$ & $ 376 $  & $122$  & $125$ \\
			\hline
				$\,\;\rho_2(?) \longrightarrow \phi(1020)\, \pi$ & $0.8$  & $0.3$  & $-$\\
			\hline
		\end{tabular}
		\caption{Decay rates of $\rho_2$ into  the vector and the pseudoscalar mesons in MeV.}\label{tab:rho2}
				\end{table}
				
We finally present the result for the well-established tensor meson $\rho_3(1690)$ in Table \ref{tab:rho3}. The decay channel of $\Gamma(\rho_3(1690)\rightarrow\omega(782) \pi) $ which is measured experimentally too, is 5-6 times smaller than $\Gamma(\rho_2\rightarrow\omega(782) \pi) $ within LQCD simulations in spite of having the same mass.

\begin{table}[h] 
		\centering
		\renewcommand{\arraystretch}{1.1}
		\begin{tabular}[c]{|c|c|c|c|}
			\hline
			Decay process (in model) & PDG \cite{Workman:2022ynf}  & eLSM   \cite{Jafarzade:2021vhh} & LQCD   \cite{Johnson:2020ilc}  \\
			\hline \hline
				$\,\;\rho_3(1690) \longrightarrow \rho(770)\, \eta$ & $-$ & $3.8\pm0.8$  & $-$  \\
			\hline
						$\,\;\rho_3(1690) \longrightarrow \bar{K}^\ast(892)\, K +\textbf{c}.\mathrm{c}.$ & $ -$  & $3.4\pm 0.7$  & $2$ \\
			\hline
		$\,\;\rho_3(1690) \longrightarrow \omega(782)\, \pi$ & $ 25.8\pm 9.8 $  & $35.8\pm 7.4$  & $22$ \\
			\hline
				$\,\;\rho_3(1690) \longrightarrow \phi(1020)\, \pi$ & $-$  & $0.036\pm 0.007$  & $-$\\
			\hline
		\end{tabular}
		\caption{Decay rates of $\rho_3(1690)$ into  the vector and the pseudoscalar mesons in MeV.}
		\label{tab:rho3}
				\end{table}

\section{Conclusion}
We have studied $\rho_2$ axial-tensor meson,  chiral partner of the tensor meson $a_2(1320)$ in the framework of a chiral model for low-energy QCD. We predict its mass to be around $1.663$ GeV similar to the Relativistic Quark model prediction. Because of the chiral symmetry, the parameter determined in the tensor sector allows to make predictions for unknown ground-state axial-tensor resonance. The effective model fitting to the LQCD results is also presented.
\section*{Acknowledgement}
I am  thankful to Adrian K{\"o}nigstein for collaboration in \cite{Jafarzade:2021vhh} ,  Milena Piotrowska, Arthur Verijeken in \cite{Jafarzade:2022uqo} and Francesco Giacosa for his reading the manuscript as well as collaboration in \cite{Jafarzade:2021vhh,Jafarzade:2022uqo}. 
I acknowledge financial support through the project ``Development Accelerator of the Jan Kochanowski University of Kielce'', co-financed by the European Union under the European Social Fund, with no. POWR.03.05. 00-00-Z212 / 18 and  support through the NCN OPUS no. 2018/29/B/ST2/02576.

\end{document}